\documentstyle[12pt]{article}
\newcommand{\be}{\begin{equation}}
\newcommand{\ee}{\end{equation}}
\newcommand{\bd}{\begin{displaymath}}
\newcommand{\ed}{\end{displaymath}}
\newcommand{\baa}{\begin{array}{lll}}
\newcommand{\eaa}{\end{array}}
\newcommand{\ba}{\begin{eqnarray}}
\newcommand{\ea}{\end{eqnarray}}
\newcommand{\la}{\label}

\newcommand{\q}{\bar q}

\begin{document}
\begin{titlepage}
\begin{flushright}
HD--THEP--95--48\\
\end{flushright}\vskip1.5cm
\begin{center}
{\bf\LARGE Integral Transform Technique}\\
\vskip.3cm
{\bf\LARGE for Meson Wave Functions}\\
\vspace{1cm}
A.~P.~Bakulev
\footnote{E-mail bakulev@thsun1.jinr.dubna.su}
and S.~V.~Mikhailov
\footnote{E-mail mikhs@thsun1.jinr.dubna.su}
 \\
\bigskip
 Bogoliubov Laboratory of Theoretical Physics,\\
  JINR, Dubna, Russia\\
and\\
Institut  f\"ur Theoretische Physik\\
Universit\"at Heidelberg\\
Philosophenweg 16, D-69120 Heidelberg
\end{center}
\vspace{2.0cm}
\begin{abstract}\noindent
In a recent paper \cite{bm_wfp} we have proposed
a new approach for extracting the wave function  of the
$\pi$-meson $\varphi_{\pi}(x)$ and the masses and
 wave functions  of its first resonances
from the new QCD sum rules for non-diagonal correlators
obtained in \cite{arnew}.  Here, we test our approach
using an exactly solvable toy model as an illustrating example.
We demonstrate the validity of the method
and suggest a pure algebraic procedure for extracting
the masses and wave functions relating to the case under
investigation.
We also explore the stability of the procedure under
perturbations of the theoretical part of the sum rule.
In application to the pion case, this results not only in
the mass and  wave function of the first resonance ($\pi'$),
but also in the estimation of $\pi''$-mass.
\end{abstract}
\end{titlepage}

\newpage

\section {Introduction}

   An important problem in the theory of strong interactions
is to calculate, from the first principles of QCD, hadronic
wave functions (WFs)
 $\varphi_{\pi}(x)$,
 $ \varphi_N(x_1,x_2,x_3), \ldots$~\cite{{CZ77}}
that  accumulate  necessary information about the
non-perturbative long-distance  dynamics of the theory.
These phenomenological functions  naturally appear as a result
of applying ``factorization theorems" to hard exclusive processes
\cite{CZ77}, \cite{ar77}, \cite{bl78}.
It seems now that only the QCD Sum Rule (SR) approach \cite{svz}
and lattice calculations \cite{lat91} can provide information
about the form of hadronic WFs.

The most popular set of hadronic wave functions, due to
~V.~L.~Chernyak and A.~R.~Zhitnits\-ky (CZ) \cite{cz82},
was produced with the help of QCD SR for the first moments
$\langle\xi^N\rangle_{h}$ ($N=2,\, 4$) of mesons WFs.
But the method they used has three main drawbacks:
\begin{enumerate}
\item reducing the initially non-local objects to a few local
ones:
$\langle \q(0) q(0)\rangle,
~~\langle  G(0)G(0)\rangle, \ldots$, --
leads to the fast growth with $N$ of the power corrections in
the  operator product expansion (see discussion in
~\cite{nlwf92});
\item the reconstruction of the whole function $\varphi_{\pi}(x)$
from just two non-trivial moments  $\langle\xi^2\rangle_{\pi}$
and $\langle\xi^4\rangle_{\pi}$  is an unreliable procedure;
\item the phenomenological model of the spectral density, based on
local quark-hadron duality, is too crude for QCD SRs for the WF.
\end{enumerate}
Now it is known that the hadronic WFs
are rather sensitive to the structure of the non-perturbative
vacuum \cite{nlwf86}.
Therefore one should use a non-local condensate
like $\langle q(0) E(0,z) q(z)\rangle$ \ and/or
 ~$\langle G(0) \tilde E(0,z) G(z)\rangle$  \cite{nlc}
which can reflect the complicated structure of the QCD vacuum.
\footnote{Here
$E(0,z)=P\exp(i \int_0^z dt_{\mu} A^a_{\mu}(t)\tau_a)$ is
the Schwinger phase factor required for gauge invariance.}

   The modified QCD SR with non-local condensates have been
constructed in \cite{nlwf86} and it has been demonstrated
that the introduction of
the correlation length for condensate distributions
produces much smaller values for the first moments of the pion WF
compared to the CZ values (see also \cite{nlwf89},~\cite{nlwf92}).
This leads to a form of the pion WF strongly different from
the CZ two-hump form and a little wider than the asymptotic
form $\varphi_{\pi}^{as}(x) = 6x(1-x)$.

In our previous paper \cite{bm_wfp} we have obtained
{\bf directly the forms} of the pion and its first resonance
WFs, using the available smooth ansatze for the correlation
functions of the non-local condensates and without any
suggestion about the quark-hadron duality for the spectral
density.
This program has been suggested and realized in \cite{arnew},
and we have developed alternative methods of extracting WFs from
this sum rule for a non-diagonal correlator ~\cite{bm_wfp}.
It should be mentioned that we do not need the principle
of stability over the Borel parameter $M^2$, and the Borel SR
appears only at an intermediate stage.

Here, we test our method using an exactly solvable model as
an  example (initiated by the two-dimensional
quantum-mechanical oscillator). We demonstrate
the validity of the method and also suggest a pure algebraic
procedure for extracting the masses and WFs of the particle
under consideration. We investigate the stability of
the procedure to perturbations of the theoretical part
of the sum rule and obtain some results to believe that it
works well in real QCD SR.
In application to the pion case, this results in producing not
only the mass and WF of the first resonance ($\pi'$)
but also to provide an estimate of the $\pi''$-mass.

Our quantitative results, i.e. the details of the WF shapes and
the values of masses, are dependent on the certain form
of the ansatz used for the non-local condensates. This form
may be obtained in a future theory of the QCD vacuum.
In the absence of such a theory the form of the ansatz was
chosen as a phenomenological ``input" (for details see
~\cite{bm_wfp}). From this point of view, our calculational
scheme may be considered as a suitable framework for connecting
the hadronic properties, on one hand, with the future
theory of the non-perturbative vacuum on the other.

\section {Non-diagonal QCD SR for pion wave functions and
the Method of Integral Transform}
\subsection {Non-diagonal QCD SR}

  The sum rule for the ``axial'' WFs
$\varphi_{\pi^{\ldots}}(x)$ of the pseudoscalar mesons,
based on the non-diagonal correlator of the axial and
pseudoscalar currents
\ba
\varphi_{\pi}(x) &+& \varphi_{\pi'}(x) e^{-m_{\pi'}^2/M^2}
+\varphi_{\pi''}(x) e^{-m_{\pi''}^2/M^2} + \ldots
\equiv  \Phi(\frac{1}{M^2},x)
\la{ar-sr} \\
&=& \frac{M^2}{2}\left (1-x+\frac{\lambda_q^2}{2M^2} \right)
f(xM^2) + (x \to 1-x), \nonumber
\ea
has been suggested in ~\cite{arnew} and possesses some
remarkable properties. For the WFs on the l.h.s.
of (\ref{ar-sr}) one has
\be
\langle \xi^N \rangle_{M} \equiv \int_0^1 \ \varphi_M(x)
(2x-1)^N \ d x; \quad
\langle \xi^{N=0} \rangle_{\pi} =1 \  , \  \  \
\langle \xi^{N=0} \rangle_{\pi'} = \langle \xi^{N=0}
\rangle_{\pi''} = \ldots =0.
\la{norm}
\ee
On the r.h.s. of (\ref{ar-sr}) there appears a single
correlation function $f(\nu)$ parametrizing the
$z^2$-dependence
\footnote{ In deriving these sum rules we can always make a
Wick rotation, i.e., we assume that all coordinates are
Euclidean, $z^2 <0$.} of the non-local quark condensate:
\ba
\langle:\bar q(0)q(z):\rangle \equiv \langle:\bar q(0)q(0):
\rangle
\int_{0}^{\infty} e^{\nu z^2/4}\, f(\nu)\, d\nu,\\
\int_{0}^{\infty}\, f(\nu)\, d\nu = 1,\
\int_{0}^{\infty}\nu f(\nu) d\nu = \frac12
\left({\langle:\bar q(0)(D^2) q(0):\rangle\over
\langle:\bar q(0)q(0):\rangle}=\lambda_q^2 \right).
 \label{eq:qq}
\ea
Here $D$ is the covariant derivative, $\lambda_q^2$ can be
treated as the average virtuality of
vacuum quarks. Equation (\ref{ar-sr}) demonstrates in the
most explicit manner that the distribution $\varphi_{\pi}(x)$
of quarks inside the pion over the longitudinal momentum
fraction $x$ on the l.h.s. of the equation is directly related
to the distribution $f(\nu)$ over the virtuality $\nu$ of the
vacuum fields on the r.h.s..

It should be noted that the sum rule (\ref{ar-sr}) results
from the approximations both in the theoretical
(for a detailed discussion, see ~\cite{arnew}) and
the phenomenological parts (see ~\cite{bm_wfp}). These are
\begin{enumerate}
\item a reduction of three-point correlators to two-point ones;
\item a reduction of the quark-quark-gluon distribution function
to a quark-quark one;
\item neglecting the contributions of higher non-local
condensates, like
$\langle \bar q G G q \rangle, \langle \bar q G G G q \rangle,
\ldots$;
\item a representation of the phenomenological spectral density
$\rho(s)$ on l.h.s. of the SR as a sum of very narrow resonances
$$\rho(s) = \delta(s)\cdot f_{\pi} \varphi_{\pi}(x)+
\sum_{i\ge 1} \delta(s-\mu_i)\cdot f_{i}\varphi_i(x).$$
\end{enumerate}

In the present work we shall use a specific form of
the correlation function $f(\nu)$, given by
\ba
\la{ansatz}
f(\nu) &=& N_q \cdot\exp\left(-\Lambda^2/\nu-\sigma\cdot\nu
\right);\\
N_q &=& {1\over 2\frac{\Lambda}{\sqrt{ \sigma}}
K_1\left(2\Lambda \sqrt{\sigma} \right)},
~~K_1(z) \mbox{ is the modified Bessel function}. \nonumber
\ea
As was established in our preceding paper \cite{bm_wfp},
this ansatz (A2-ansatz there) conforms with pion physics.
A similar form for the ansatz also appears in the framework of
the instanton model for the non-perturbative vacuum
\footnote{we are grateful to A.~Dorokhov for informing us about
this result (private communication)}. Here, the parameter
$\Lambda^2 \approx 0.2\,\mbox{GeV}^2$ was extracted
from the meson QCD SR for the heavy-quark effective theory
~\cite {arnew}; the normalization constant $N_q$ and
the parameter $\sigma$ are fixed by Eq.(\ref{eq:qq}).
For the average virtuality of vacuum quarks we take
the usual QCD sum rule value
$\lambda_q^2 \simeq 0.4\,\mbox{GeV}^2$ \ \cite{belioffe}.

  As is clear from the structure of the r.h.s. of
(\ref{ar-sr}), the form of the function
$\Phi(\frac{1}{M^2},x)$ on the variable $x$ is given by two
humps (one centered at $x_A=s_A/M^2$, where $s_A$ is
the point of maximum for the ansatz correlation function
$f(\nu)$, and the other  at $x_A=1-s_A/M^2$) moving as $M^2$
changes. When $M^2$ increases, the humps become narrower,
higher and more close to the boundary points $x=0$ or $x=1$.
However, $\Phi(\frac{1}{M^2},x)$ is not yet the pion WF:
the larger $M^2$, the larger the contamination from higher
pseudoscalar states $\pi'$,  $\pi'' , \ldots$.
Due to the properties summarized in (\ref{norm}),
the corresponding WFs should oscillate
(see, e.g., Fig. \ref{fig-A2-res}); therefore the function
$\Phi(\frac{1}{M^2},x)$ does not resemble to $\varphi_\pi(x)$
at not sufficiently low $M^2$.

At low $M^2$, the pion WF dominates in the total sum
$\Phi(\frac{1}{M^2},x)$ (however, one cannot take too low $M^2$
because the operator product expansion fails for
$M^2 < \lambda_q^2$). When $M^2=0.4\,~GeV^2$, it was observed
that $\Phi(\frac1{M^2},x)$ is very close to the asymptotic
wave function of the pion (see \cite{arnew}).
 But there is no strict criterion for selecting the value of
 $M^2$ to determine the WF of the ground state (pion). Our
 method of extracting the WFs and masses of resonances from
 (\ref{ar-sr}) allows one to avoid this problem.

\subsection {The Method of Integral Transform}

  In what follows it is convenient to use a new variable
$\tau \equiv 1/M^2$ instead of the Borel parameter $M^2$.
The r.h.s. of SR (\ref{ar-sr}), i.e., $\Phi(\tau,x)$ is then
defined for $\tau \in \big[0, \frac{1}{\lambda_q^2}\big]$.
Let us consider this function in the whole complex plane
of $\tau$ and define
the integral transformation (IT) $P(N, \tau_0, \omega)$:
\ba   \la{2}
P(N, \tau_0, \omega)\big(\cdots\big) = \frac{N!}{2\pi i
\omega^N} \int_{C} \frac{\exp(\omega (\tau-\tau_0))}
{(\tau-\tau_0)^{N+1}} \big(\cdots\big) d\tau~,
\ea
where the integration is performed along the
 vertical line $C=(c-i\infty,~c+i\infty)$ with $c$
lying to the right with respect to any pole of the integrand,
~~$\omega > 0$ and $\tau_0\in(0,\frac{1}{\lambda_q^2}\big]$.
\begin{figure}[ht]
\unitlength=1.60pt
\special{em:linewidth 0.8pt}
\linethickness{0.8pt}
\begin{picture}(287.00,134.00)
\put(21.00,124.00){\vector(0,1){7.00}}
\put(94.00,74.00){\vector(0,1){15.00}}
\put(108.90,119.24){\circle{14.00}}
\put(108.90,119.24){\makebox(0,0)[cc]{$\tau$}}
\put(90.90,63.10){\makebox(0,0)[cc]{$c$}}
\put(122.00,73.00){\makebox(0,0)[cc]{Re($\tau$)}}
\put(34.81,129.00){\makebox(0,0)[cc]{Im($\tau$)}}
\put(15.95,63.10){\makebox(0,0)[cc]{$0$}}
\put(50.19,68.14){\makebox(0,0)[cc]{$\times$}}
\put(31.76,92.14){\makebox(0,0)[cc]{$\times$}}
\put(31.76,43.19){\makebox(0,0)[cc]{$\times$}}
\put(100.00,17.00){\makebox(0,0)[cc]{$C$}}
\put(121.00,68.00){\vector(1,0){7.00}}
\put(70.00,68.00){\makebox(0,0)[cc]{$\bullet$}}
\put(21.00,68.00){\makebox(0,0)[cc]{$\bullet$}}
\put(50.00,76.00){\makebox(0,0)[cc]{$\tau_0$}}
\put(69.00,62.00){\makebox(0,0)[cc]{$\lambda_q^{-2}$}}
\put(200.00,124.00){\vector(0,1){7.00}}
\put(249.00,74.00){\vector(0,1){15.00}}
\put(263.90,119.24){\circle{14.00}}
\put(263.90,119.24){\makebox(0,0)[cc]{$\tau$}}
\put(245.90,63.10){\makebox(0,0)[cc]{$c$}}
\put(277.00,73.00){\makebox(0,0)[cc]{Re($\tau$)}}
\put(213.81,129.00){\makebox(0,0)[cc]{Im($\tau$)}}
\put(194.95,63.10){\makebox(0,0)[cc]{$0$}}
\put(175.19,68.14){\makebox(0,0)[cc]{$\times$}}
\put(175.76,86.14){\makebox(0,0)[cc]{$\times$}}
\put(175.76,32.19){\makebox(0,0)[cc]{$\times$}}
\put(255.00,17.00){\makebox(0,0)[cc]{$C$}}
\put(276.00,68.00){\vector(1,0){7.00}}
\put(200.00,68.00){\makebox(0,0)[cc]{$\bullet$}}
\put(209.00,76.00){\makebox(0,0)[cc]{$\tau_0$}}
\put(167.00,63.00){\makebox(0,0)[cc]{$\lambda(x)$}}
\put(53.00,2.00){\makebox(0,0)[cc]{A -- General case}}
\put(216.00,2.00){\makebox(0,0)[cc]{B -- Toy model case}}
\put(209.19,68.14){\makebox(0,0)[cc]{$\times$}}
\put(175.76,104.14){\makebox(0,0)[cc]{$\times$}}
\put(175.76,50.14){\makebox(0,0)[cc]{$\times$}}
\put(175.76,122.14){\makebox(0,0)[cc]{$\times$}}
\put(175.76,14.14){\makebox(0,0)[cc]{$\times$}}
\put(132.00,68.00){\line(-1,0){127.00}}
\put(160.00,68.00){\line(1,0){127.00}}
\put(21.00,134.00){\line(0,-1){126.00}}
\put(94.00,8.00){\line(0,1){126.00}}
\put(200.00,134.00){\line(0,-1){126.00}}
\put(249.00,134.00){\line(0,-1){126.00}}
\end{picture}
\caption{Here $C$ is the contour of integration in the complex
plane of $\tau$; crosses ($\times$) denote the positions
(in case A -- possible) of the integrand poles.}
\label{fig-contour}
\end{figure}
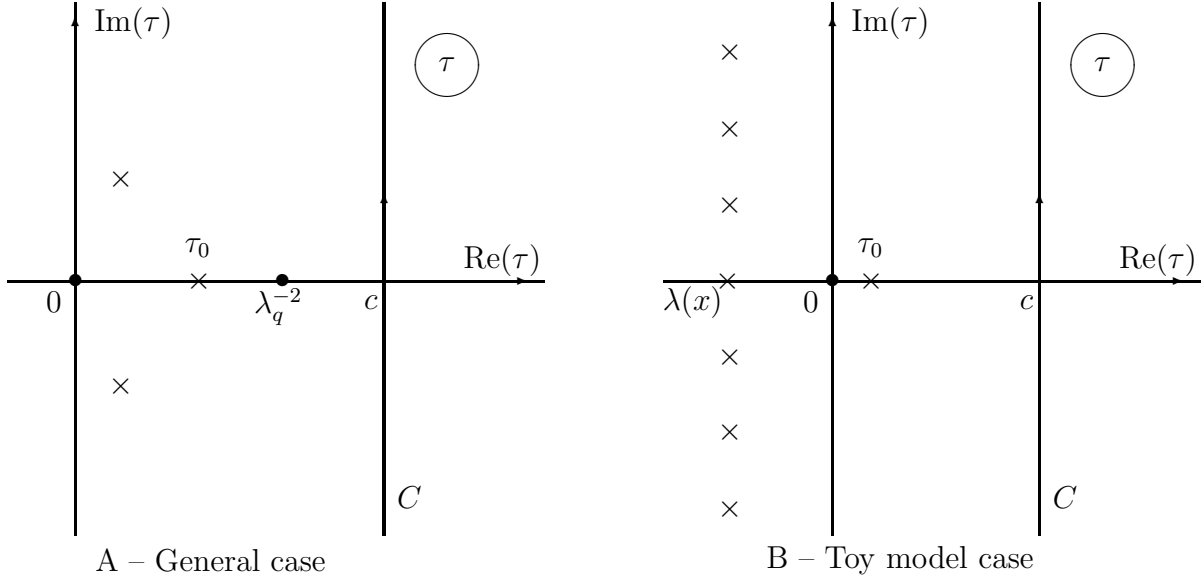

To obtain the result of this operator action on the initial
SR-representation (here $\mu_i=m_i^2$)
\be \la{3}
 \varphi(x)+ \sum_{i=1} \varphi_i(x) \exp(-\tau \mu_i)
 = \Phi(\tau,x),
\ee
consider its action on a simple exponential $e^{-\tau\mu}$.
Evidently, if $\omega>\mu$ we can close the contour $C$ to
the left and get (due to the well-known residue theorem)
the contribution from the pole at $\tau=\tau_0$; if $\mu>\omega$
we can close the contour to the right and, due to the absence
of poles in that part of complex $\tau$-plane, obtain $0$;
therefore
\be
\la{4}
P(N, \tau_0, \omega) \exp(-\mu \tau) = \theta(\omega-\mu) \cdot
\exp(-\mu\tau_0) \left(1-\frac{\mu}{\omega}\right)^N.
\ee
Then, for $\omega > 0$, we have
\be \la{eq:step}
\varphi(x)+ \sum_{i\ge 1}\theta\left(\omega-\mu_i\right)
\varphi_i(x)\left(1-\frac{\mu_i}{\omega}\right)^N
\exp\left(-\tau_0\mu_i\right)=P(N,\tau_0,\omega) \Phi(\tau, x)
\equiv \Phi_N(\omega,x).
\ee
We see that by varying $\omega$ one can switch on (switch off)
 successively  more and more resonances on the l.h.s. of the SR.
So, we have obtained a real step division (smooth at $N>0$)
of the set
of resonances by the parameter $\omega$ in (\ref{eq:step}),
instead of the exponential suppression of the whole set of
resonances with respect to the inverse Borel parameter $\tau$
in (\ref{3}). Accordingly, the hadronic characteristics
$\{\mu_i, \varphi_i(x)\}$ are determined by the positions
of singularities ( and zeros ) of the correlation function
$f(1/\tau)$ in the complex $\tau$-plane.

Let us assume now that the mass values are known
$\mu_0 =0,~ \mu_1=\mu_{exp}\approx 1.7 \mbox{GeV}^2,~ ... $.
Then, we can obtain an expression for $\varphi(x)$ by employing
equation (\ref{eq:step}) with the parameter $\omega=\omega_1
\leq \mu_1$ that corresponds to the saturation of the ground
state $\varphi(x)$
\be \la{9}
\varphi(x)=P(N,\tau_0,\omega_1)\Phi(\tau,x)=\Phi_N(\omega_1,x).
\ee
The final result should not depend on the parameter $N$
strongly. However, in a real life we don't know the position
$\mu_1$ exactly; therefore we may get  contamination from the
next state. This contamination becomes smaller when $N$
increases due to the power suppression near the threshold
$\mu_1$ (see l.h.s. of Eq. (\ref{eq:step})).
One should also take into account that the WF form
is saturated smoothly in the region $\omega < \mu_1$ and
its norm reaches unity at the end of the region. The curves
corresponding to the function
$\varphi(x)=\Phi_N(\omega_1 \approx \mu_1,x)$ are shown in
Fig. \ref{fig-A2-123} for different values of $N$. Their form
varies very slightly with $N=1,\ 2, \ 3$ (the same happens
when the parameter $\tau_0$ decreases from $2.5$ to
$1.5 \ \mbox{GeV}^2$, see the discussion in ~\cite{bm_wfp})
and is always somewhat narrower than the asymptotic one. Note
that formula (\ref{9}) may be easily generalized to a state
with a finite width  (see ~\cite{bm_wfp}). The contribution
to the pion WF due to finite width effects is not significant
numerically.

  We now describe the pure algebraic procedure that allows
one to extract from the SR (\ref{eq:step}) both the positions
($\mu_i$) and the WFs ($\varphi_i(x)$).
The procedure will be based on the simple step representation
appearing on the l.h.s. of (\ref{eq:step}).
Indeed, let us consider the difference
\be
D_1\Phi_N(\omega,x)\equiv\Phi_N(\omega,x)-\Phi_{N+1}(\omega,x).
\ee
We see that the same contributions due to $\varphi(x)$ are
cancelled out and this difference becomes non-zero just after
$\omega > \mu_1$:
\be \la{eq:D1}
D_1\Phi_N(\omega,x) =
\sum_{i\ge 1}\theta\left(\omega-\mu_i\right)\varphi_i(x)
\exp\left(-\tau_0\mu_i\right)\frac{\mu_i}{\omega}
\left(1-\frac{\mu_i}{\omega}\right)^N.
\ee
Therefore, this difference is a trigger for the first resonance
-- the function $D_1\Phi_N(\omega,x)$ must have a root at
$\omega=\mu_1$ that fixes the resonance (at fixed $x$).
Analogous trigger-like differences could be constructed for
higher resonances. By induction, let the difference $D_n$ is
trigger-like for the $n$-th resonance (i.e., it has a root at
$\omega=\mu_n$) and has the following form
\be \la{eq:Dn}
D_n\Phi_N(\omega,x) =
\sum_{i\ge n}\theta\left(\omega-\mu_i\right)\varphi_i(x)
\exp\left(-\tau_0\mu_i\right)\
\frac{\mu_i}{\omega}
\frac{\mu_i-\mu_1}{\omega}\ldots\frac{\mu_i-\mu_{n-1}}{\omega}
\left(1-\frac{\mu_i}{\omega}\right)^N.
\ee
Then the difference $D_{n+1}$ is defined recursively by
the recurrence relation
\be \la{eq:Dn+1}
D_{n+1}\Phi_N(\omega,x) =\left(1-\frac{\mu_n}{\omega}\right)
D_n\Phi_N(\omega,x) - D_n\Phi_{N+1}(\omega,x).
\ee
For instance, for the second resonance the corresponding
difference is of the form
\be \la{eq:D2}
D_2\Phi_N(\omega,x) =\left(1-\frac{\mu_1}{\omega}\right)
D_1\Phi_N(\omega,x) - D_1\Phi_{N+1}(\omega,x),
\ee
and it yields the value of $\mu_2$ provided the value of
$\mu_1$ is known from the previous step. Hence, with the help
of these trigger-like differences one can easily determine
successively  the masses $\mu_i$. After that it is
straightforward to determine the WFs $\varphi_i(x)$ itself,
i.e.:
\be \la{phi1}
\varphi_1(x) = \frac{\left( \Phi_1(\omega,x) -
\Phi_1(\mu_1,x)\right) \exp\left(\tau_0\mu_1\right) }
{ \left(1-\frac{\mu_1}{\omega}\right) }
\ee
where the value of the parameter $\omega \in
\left(\mu_1,\mu_2\right]$
corresponds to saturation of the first
resonance. One can try the other formula for $\varphi_1(x)$
following from (\ref{eq:D1}), viz.
\be \la{phi1N} \varphi_1(x) =
\frac{D_1\Phi_N(\omega,x)\exp\left(\tau_0\mu_1\right)}
{\frac{\mu_1}{\omega} \left(1-\frac{\mu_1}{\omega}\right)^N}.
\ee
 The function $\varphi_1(x)$ in (\ref{phi1N}) should not
depend on the value of $N$ in an appreciable manner
when the saturation of the resonance is reached.

\section {An exactly solvable toy model}

The above formulated procedure is based on the understanding of
the initial SR (\ref{ar-sr}) as an exact equality. But in the
real case the connection between the non-perturbative vacuum
distribution $f(\nu)$ on the r.h.s of SR and the pseudoscalar
meson properties reflected on by the spectral expansion
$\sum_i\varphi_i(x)\exp(-\mu_i\tau)$ on the r.h.s. of SR is
only {\bf an approximation}.
To understand the reliability of the suggested technique (IT)
as an approximated equality, it should be useful to check it,
first, with a model $\Phi(\tau,x)$ having an exact spectral
expansion. After that we deform $\Phi(\tau,x)$ in some special
way to destroy the exact equality in the spectral expansion with
the initial WFs and masses, and apply our procedure again.
Following this way, one may investigate the stability of the
procedure to perturbations and its suitability.
We hope, despite the fact that the exact equality is broken,
that the procedure will ``work" and that the new extracted
parameters $\{ \mu_i', \varphi_i'(x)\}$ will be close to
the initial values, if the  is not too strong.
Only in this case our procedure makes sense.

\subsection {Formulating the model}

We should like to test our approach for some simple exactly
solvable model. We select one which originates from the 2D
harmonic oscillator problem in quantum mechanics (for more
details see ref.\cite{vzns-qm}). The Borel transformed Green
function $M(\tau,x)$ of the 2D harmonic oscillator
(with $\omega_0=1$) in the coordinate representation
\be
M(\tau,\mbox{\bf r}) \equiv \sum_{k=0}|
\psi_k(\mbox{\bf r})|^2e^{-\tau E_k}
\ee
has for $\mbox{\bf r}=0$ the form
\be
M(\tau,0) = \frac{m}{2\pi \sinh(\tau)}.
\ee
The spectrum of this model (as one can easily see from the
geometric progression summation formula) is equidistant with
a two-fold step
\be
 E_k = 2 k + 1.
\ee
We can also write down an explicit analytical form for
$M(\tau,\mbox{\bf r})$ with $\mbox{\bf r}\neq 0$, but
there is no need for this, since we shall work, for simplicity,
with a modified quantity. To this end, we define the function
\be
\Phi(\tau,x;\phi) = \frac{1}{1 - \phi(x) \exp(-2\tau)}
\ee
which obviously has the spectral expansion we are interested
in, viz.
\be
\Phi(\tau,x;\phi) = \sum_{k=0}\phi(x)^ke^{-2\tau k}.
\ee
This toy model has a nice feature: all its resonance WFs are
defined through one WF, namely the WF of the first resonance
$\varphi_1(x)=\phi(x)$ which can be selected at free choice.
For our convenience, we take the following form
\be
\phi(x) = 4x(1-x),
\ee
i.e., the asymptotic WF of the pion
(albeit with wrong normalization).

\subsection {IT-technique for the toy model}

  For evaluating $\Phi_N(\omega,x)$ in our toy model, we use
the well-known residues-technique. Separating residues
(see Fig. \ref{fig-contour}) from the $\tau_0$-pole
($\Delta_{\tau_0}\Phi_N$) and the poles
$\tau_n(x) \equiv \lambda(x) + i \pi n$ with
$\lambda(x) \equiv \frac{1}{2} \log(\phi(x))$
($\Delta_{\Phi}\Phi_N$), we obtain
\ba
\la{rhs-toy}
\Phi_N(\omega,x) &\equiv& \Delta_{\tau_0}\Phi_N +
\Delta_{\Phi}\Phi_N\\
\Delta_{\tau_0}\Phi_N &=&
\theta(\omega)\frac{\exp(-\omega\tau_0)}{\omega^N}
\frac{d^{(N)}}{d\tau_0^N}
\left[\frac{\exp(\omega\tau_0)}
{1-\varphi(x)\exp(-2\tau_0)}\right]\\
\Delta_{\Phi}\Phi_N &=& - \theta(\omega)\Gamma(N+1)
\frac{\exp\left\{-\omega\left[\tau_0-\lambda(x)\right]\right\}}
{(-\omega)^N}\times\nonumber\\
&\times& \left[\frac1{2(\tau_0-\lambda(x))^{N+1}}
+\sum_{n\ge1}\frac{\cos\left[\pi n\omega+(N+1)\phi_n\right]}
   {\left[\left(\tau_0-\lambda(x)\right)^2+\pi^2n^2\right]
    ^{(N+1)/2}}\right].
\ea
Here
$\phi_n\equiv\arctan
\left[\pi n/\left(\tau_0-\lambda(x)\right)\right]$.

  The l.h.s. of SR (\ref{eq:step}) in this model has the form
\be
\la{lhs-toy}
1+ \sum_{k\ge 1}\theta\left(\omega-2k\right)\phi(x)^k
\left(1-\frac{2k}{\omega}\right)^N\exp\left(-2k\tau_0\right).
\ee
By this way we obtain the exact SR in the form
\begin{center}
 (\ref{rhs-toy}) $=$ (\ref{lhs-toy}).\\
\end{center}
Saying ``exact" we mean, that this SR is valid for all
$x\in [0,1]$ and for all $\omega$. Moreover, it holds also
for all choices of the initial WF $\phi(x)$!
This behavior is shown explicitly in Fig. \ref{fig-step},
where the step-like structure on the l.h.s. of SR
(see (\ref{lhs-toy})) is reproduced by its r.h.s.
(see (\ref{rhs-toy})). Our trigger-like differences exactly
determine masses $\mu_k = 2 k$ and the corresponding WFs
$\varphi_k(x)=\phi(x)^k$.

\subsection {Deformations of $\Phi(\tau,x)$ and the IT-procedure}

All formulas of the preceding subsection are exact
(they are in some sense identities).
But in real QCD problems we have no such regular behavior of
the theoretical part of the SR and one can use the spectral
expansion only approximately. Moreover, we
cannot be sure about the adequacy of ansatz (\ref{ansatz})
for the true non-perturbative amplitude in the whole region of
variables. Having this in mind, let us inspect the stability
of the suggested procedure under perturbations of the r.h.s.
of (\ref{eq:step}).

The first problem here is the following: how can one deform
the original function $\Phi(\tau,x)$ while retaining the
possibility of controlling the degree of this deformation?

  We apply the following method. As we know, the function
$\Phi(\tau,x)$ can be expanded in a series of poles over
the variable $\tau$:
\be
\Phi(\tau,x;\phi) \equiv \frac{1}{1-\phi(x)\exp(-2\tau)} =
\frac{1}{2}+\sum_{n=-\infty}^{\infty}\frac{1}{2(\tau-\tau_n(x))}.
\ee
 If we exclude in this expansion a set of poles, we obtain some
deformed function. It should be stressed that this
function does not have the same spectral expansion as
the original one.
Let us consider our IT-procedure with these deformed functions.
The exact result for the difference $D_1\Phi_1$ is reproduced
in Fig. \ref{fig-D1_All}. The deformations
we have determined {\bf will destroy} the exact step-structure on
the l.h.s. of (\ref{eq:step}) and (\ref{eq:Dn}). This procedure
proceeds as follows.
\begin{enumerate}
\item
At first we exclude all the poles $\tau_n(x)$ with $|n| > K$
(the corresponding deformed $\Phi$ will be denoted as
$\Phi(\omega,x;K)$). Fig. \ref{fig-D1_K6} contains the curve
for $D_1\Phi_1(\omega,x;K=6)$. As we can conclude in this case,
the situation is fine:
the main deviations from $\theta(\omega-2)$ are located
near the origin ($\omega\sim 0.2$).
The locations of resonances are well-determined.
\item
At the next step we exclude also the last six poles
($|n|=3, 4 \, \mbox{and}\, 5$).
The corresponding curve for $D_1\Phi_1(\omega,x;K=3)$
is reproduced in Fig. \ref{fig-D1_K3}. The situation is
still good:
though deviations from $\theta(\omega-2)$ are larger
relative to $K=6$, they do not extend over $\omega\sim 0.2$.
The locations of resonances are also well-determined
and our ``triggering'' is still valid.
\item
The last step is to exclude the same poles as in the first
case, but in addition also those with $n=\pm 1$. The curve
for $D_1\Phi_1(\omega,x;K=6+1)$ is reproduced in
Fig. \ref{fig-D1_K61}. Now the pattern is changed:
deviations form the initial form are large and they reach
$\omega = 1$. Indeed, this new form is rather close to
$\theta(\omega-1)$: there exists the zero for $\omega = 1$
where the original function has no resonances at all.
Going through the procedure with this value ($\mu_1=1$)
for the mass of the first resonance, we obtain the new
spectrum $\mu_k=k$ with the new resonance WFs
$\left\{\varphi_k(x)\right\}\not=\left\{\phi^k(x)\right\}$.
The crucial indication for the correctness of such an
interpretation is the smallness of the deficit of the initial SR
\ba
\Delta_{SR}\{\Phi(\tau,x;K=6+1)\}
&\equiv& \Phi(\tau,x;K=6+1)\\
&&-\left[\varphi_0(x)+
\varphi_1(x)\exp(-\mu_1\tau)+\varphi_2(x)\exp(-\mu_2\tau)\right]
+\ldots .
\nonumber
\ea
We evaluate this quantity for the new and for the initial spectra
and obtain a clear signal that the new one provides the genuine
description of the SR (see Fig. \ref{Def-SR-K=61_new} and
\ref{Def-SR-K=61_old}).
\end{enumerate}

As to the extracted wave functions  $\varphi_k(x)$,
we can characterize their evolution under deforma\-tions
by the following statements:
\begin{itemize}
\item
opposed to the exact solution, $\left\{\varphi_k(x)\right\}$ become
$\omega$-dependent when $\omega\in(\mu_i;\mu_{i+1}]$.
They tend to the exact ones with $\omega$ more close to
the right end of the corresponding interval
(see Fig. \ref{fig-WF_res12});
\item
the difference between $\varphi_k(x)$ and the exact WF of
the $k$-th resonance becomes stronger for higher $k$ (i.e.
for highly exited states; see Fig. \ref{fig-WF_res12}).
\end{itemize}

To conclude these investigations:
the method really works, even for strong deformations
of the structure of the initial spectrum as the last example
of deformation demonstrates. Moreover, we see
that deformations of the analytical structure
of the initial $\Phi(\tau,x;\phi)$ in the complex $\tau$-plane
far from (near) the main pole $\tau=\lambda(x)$ produce small
(large) deformations of the spectrum and the WFs.
Therefore, the knowledge of the analytic properties
of the function $\Phi(\tau,x)$ in a region in the complex
$\tau$-plane when includes the pole $\tau=\lambda(x)$\ and
a few others appears to be sufficient for obtaining
information on the first few resonances and their WFs.
We hope that an analogous situation appears in the SR for pion.

\section {IT-procedure for the pion and its resonances}

\begin{figure}[h]
  \caption{Form of the WF of the first resonance
  $\varphi_{\pi'}(x)$ extracted from
  $D_1\Phi_1(\omega\approx 3~\mbox{GeV}^2,x)$ for the pion case.}
     \label{fig-A2-res}
      \end{figure}
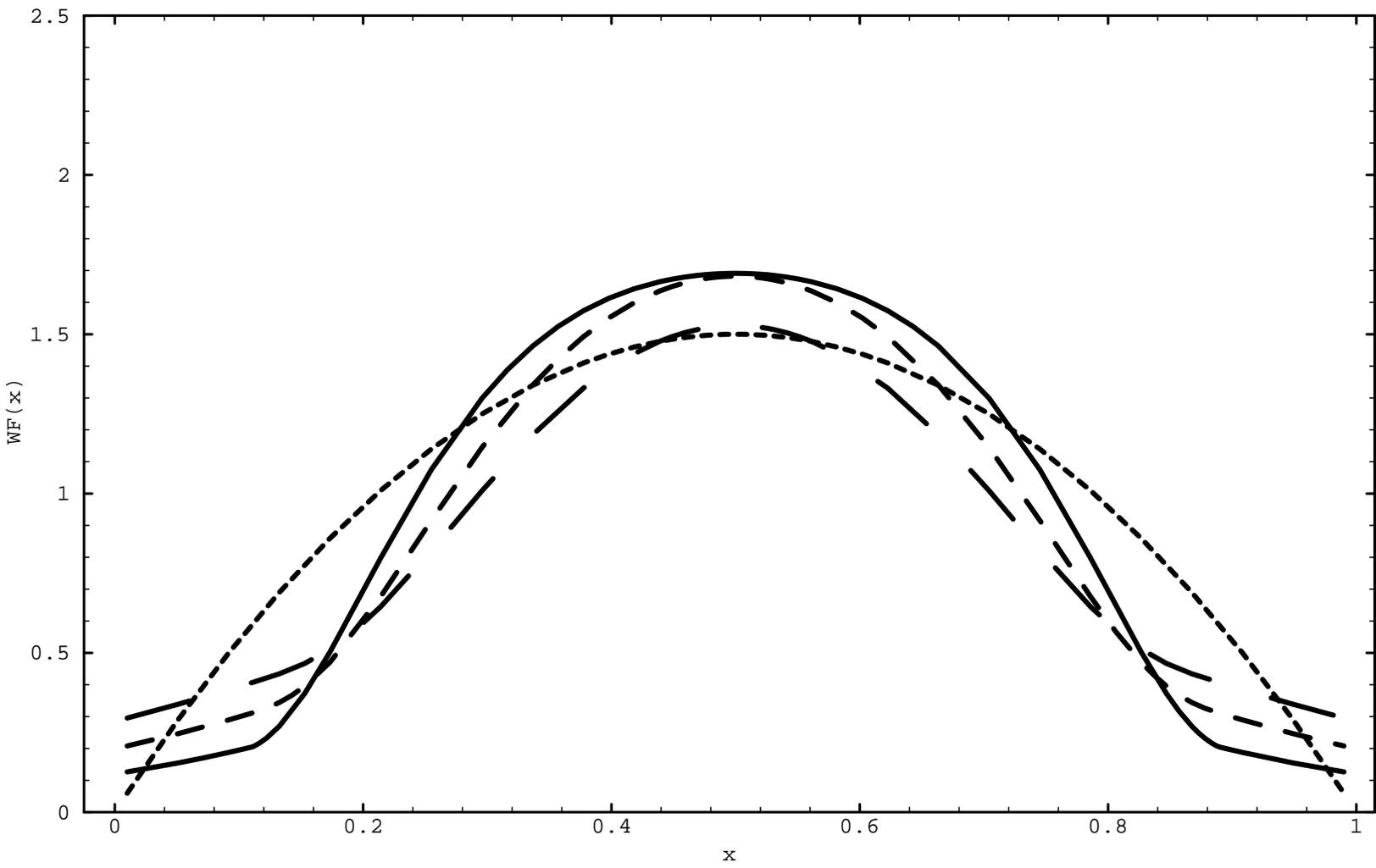
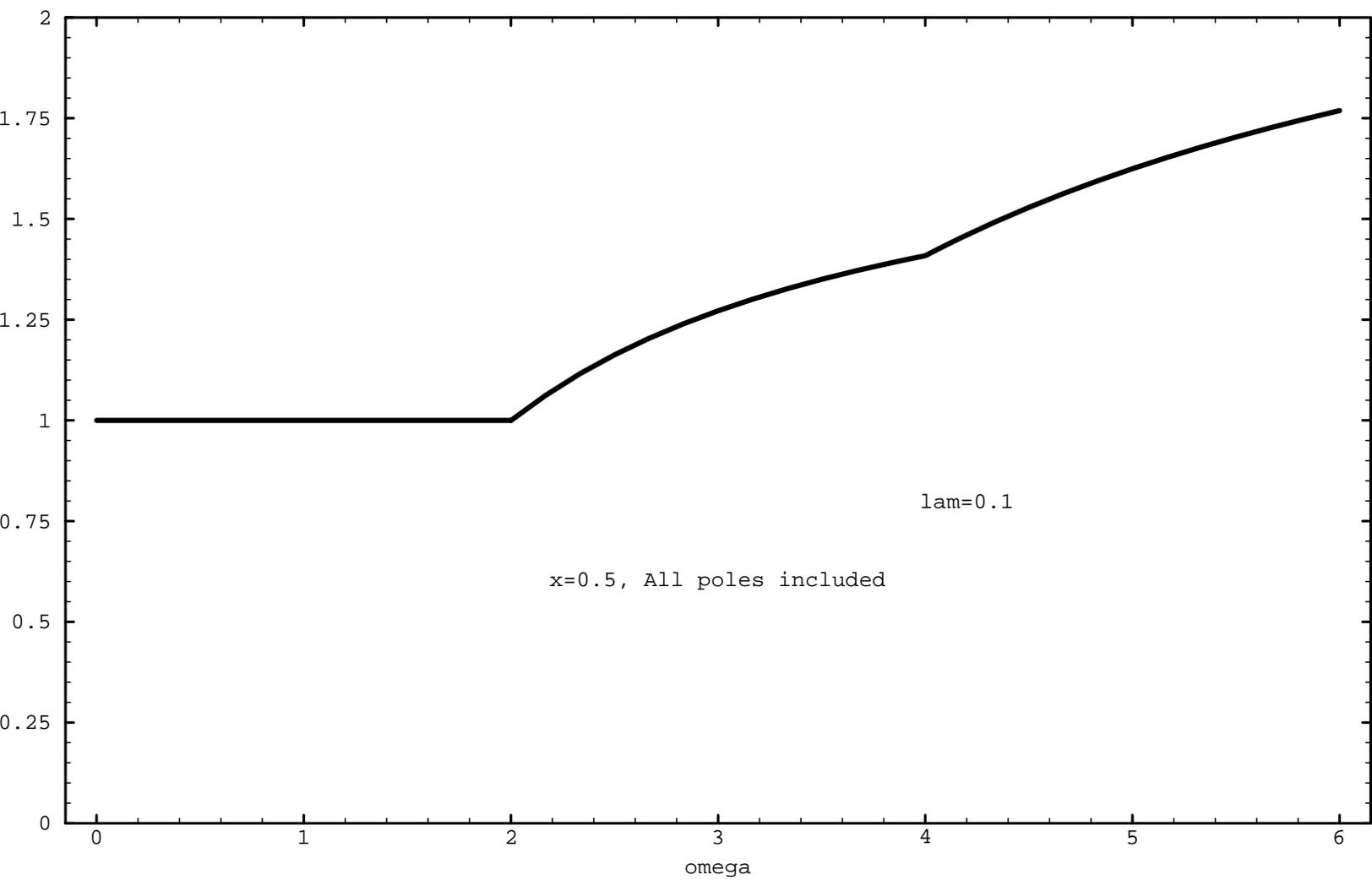
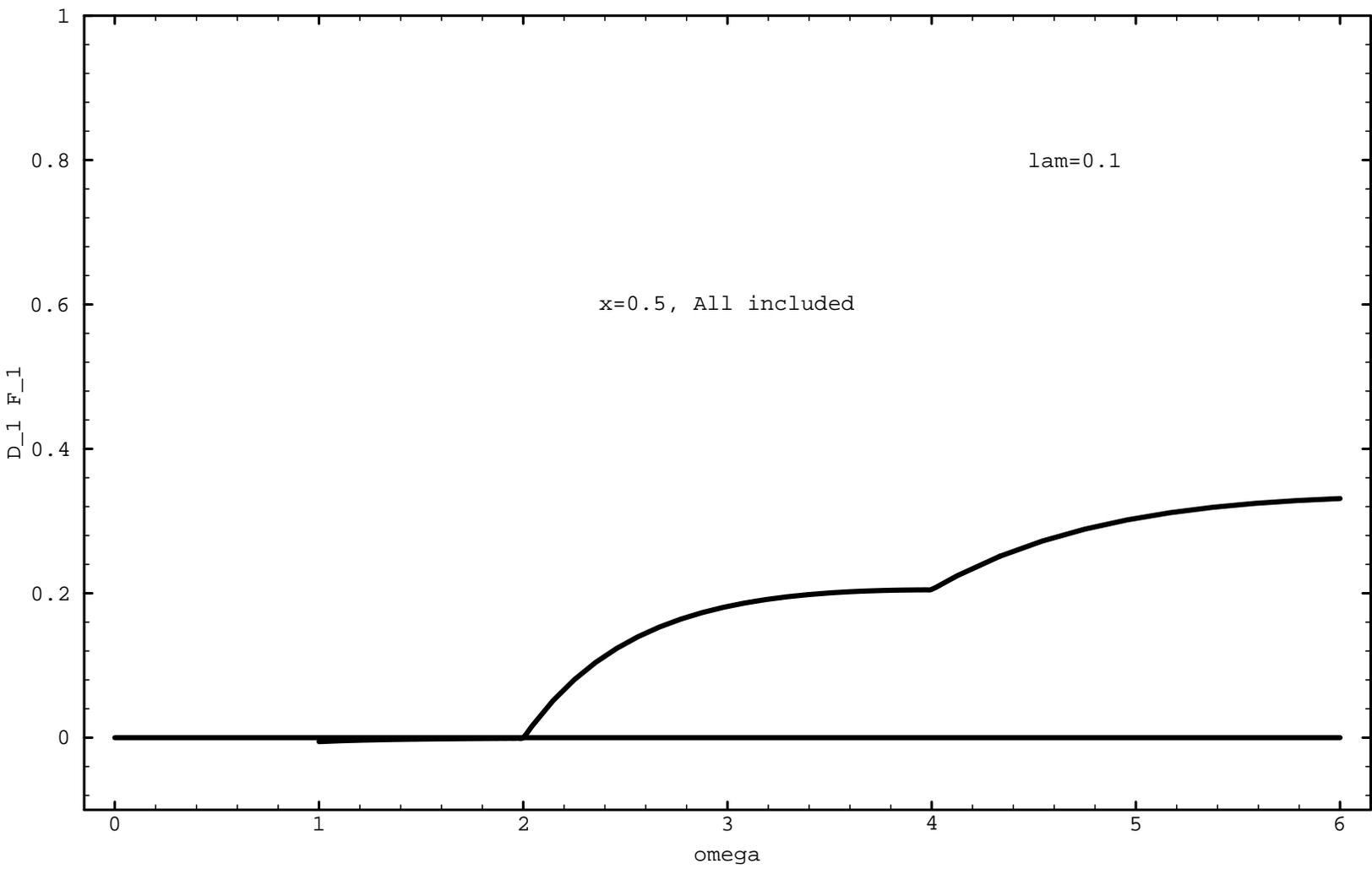
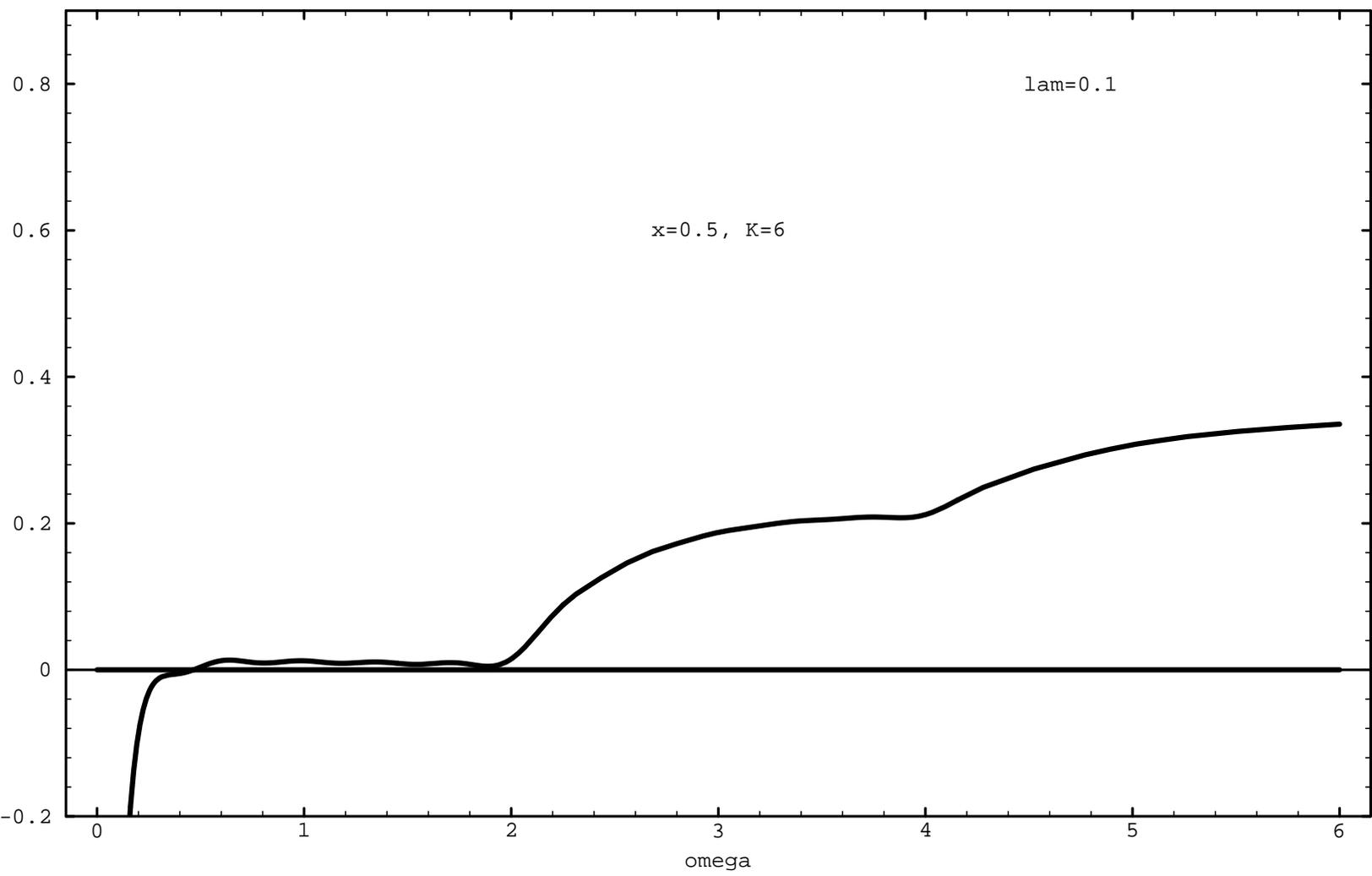
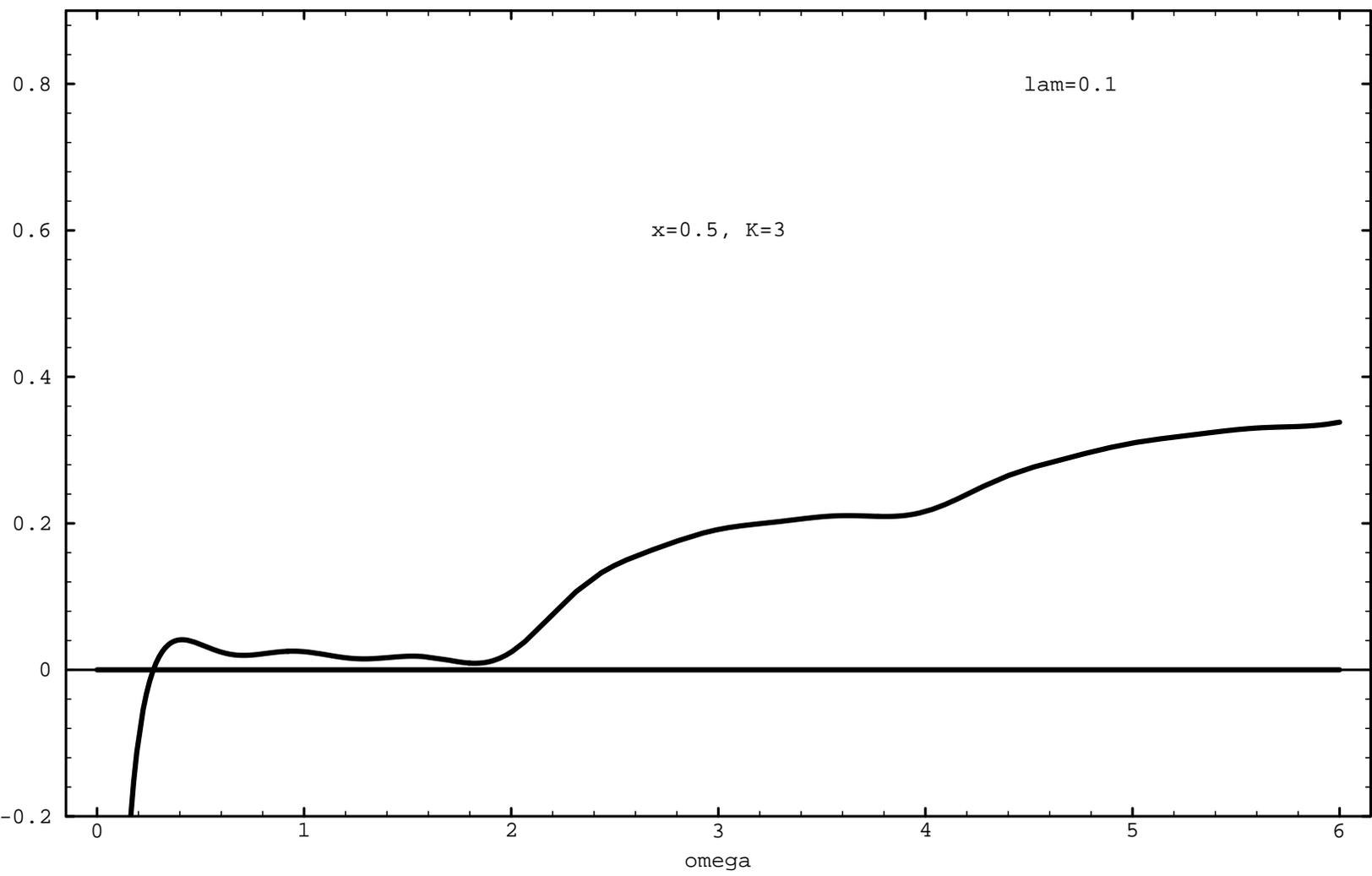
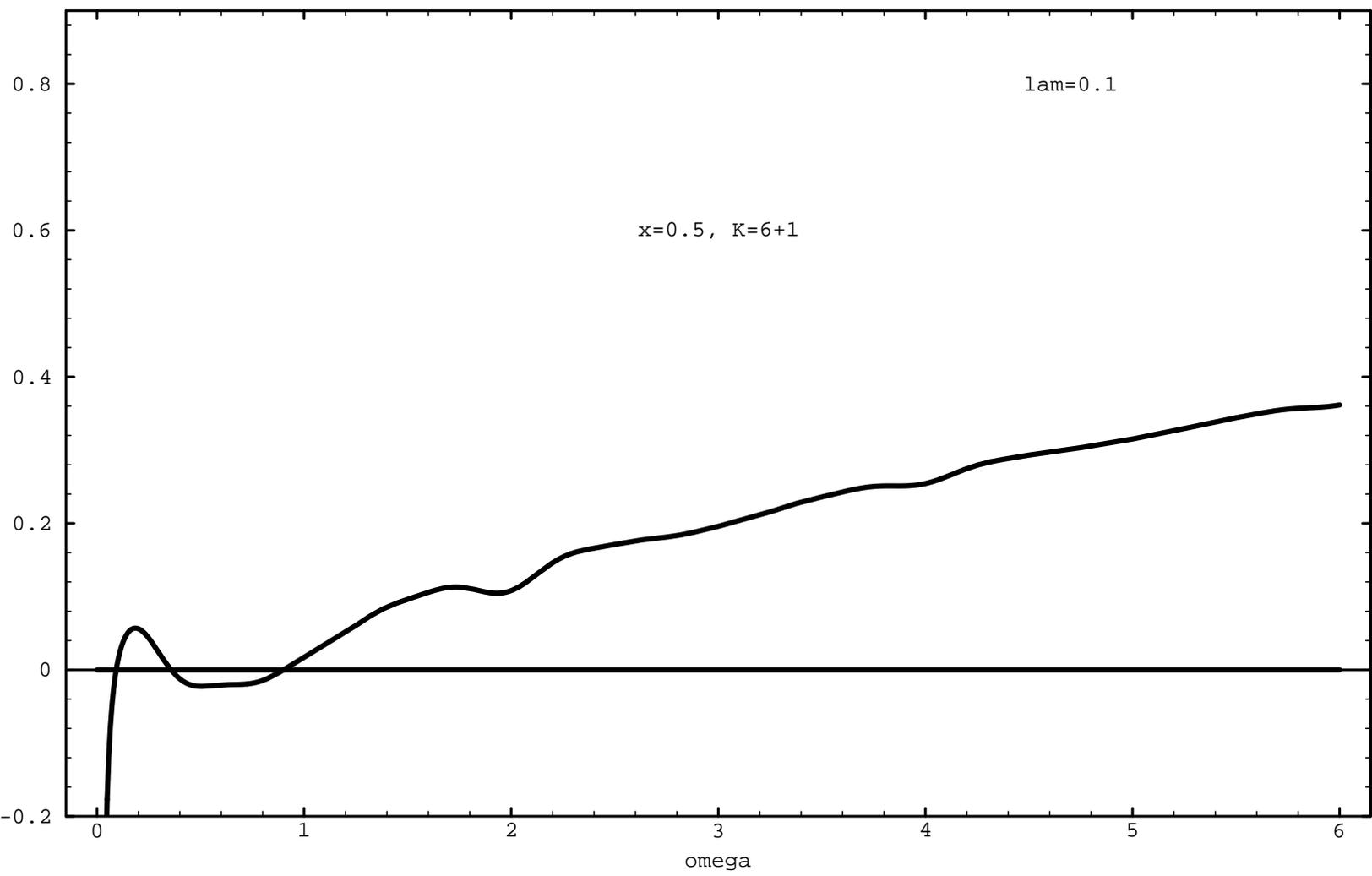
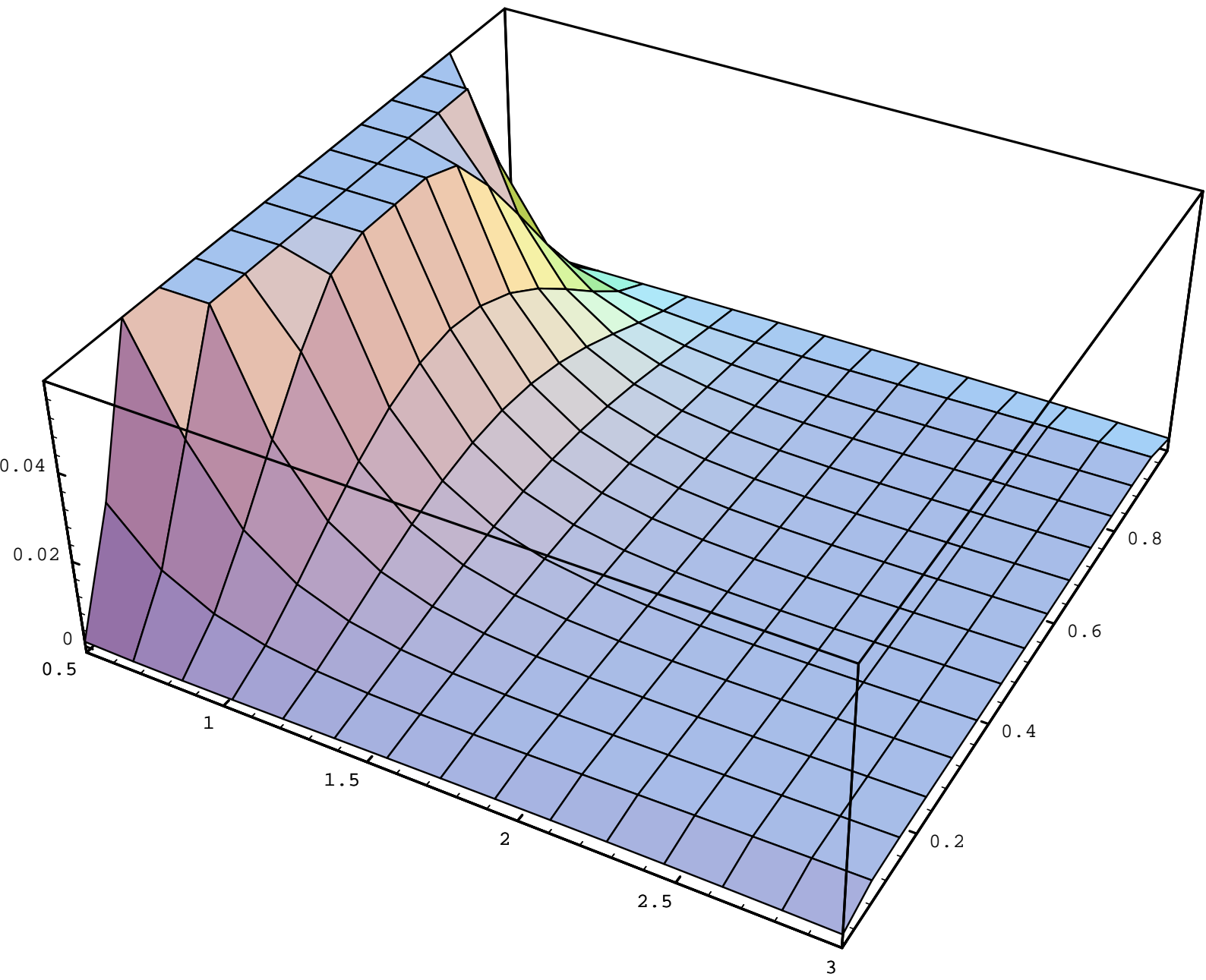
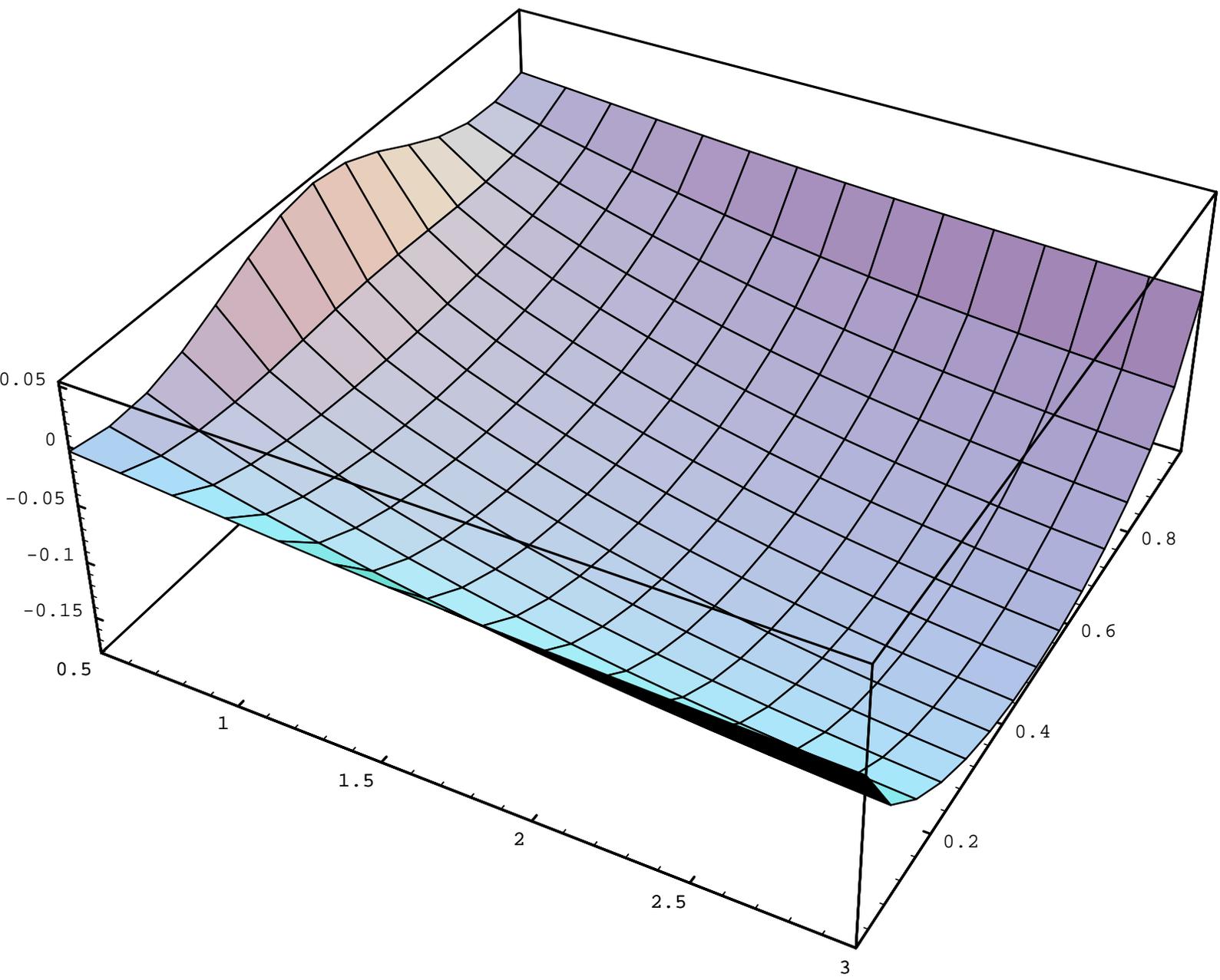
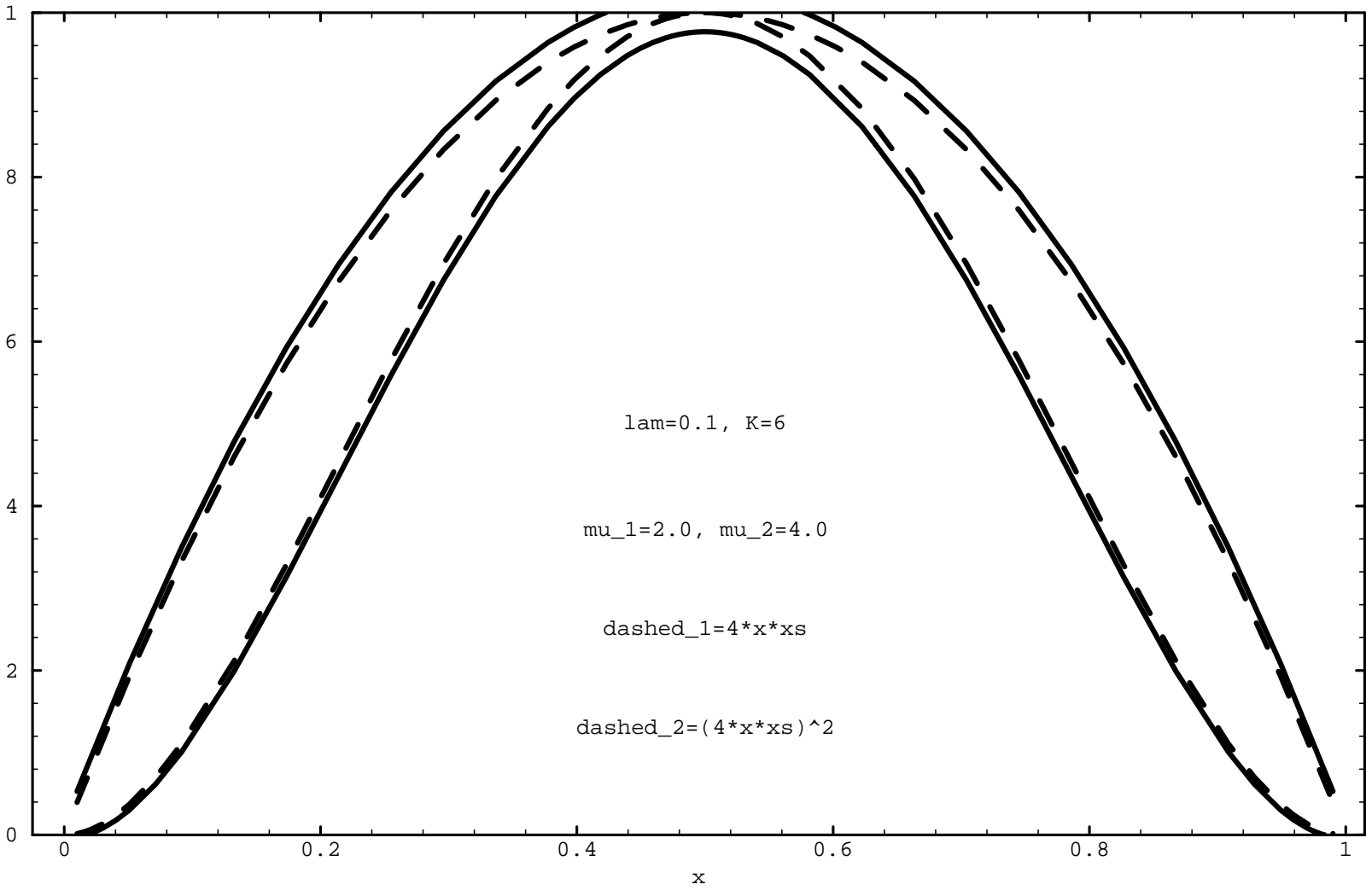
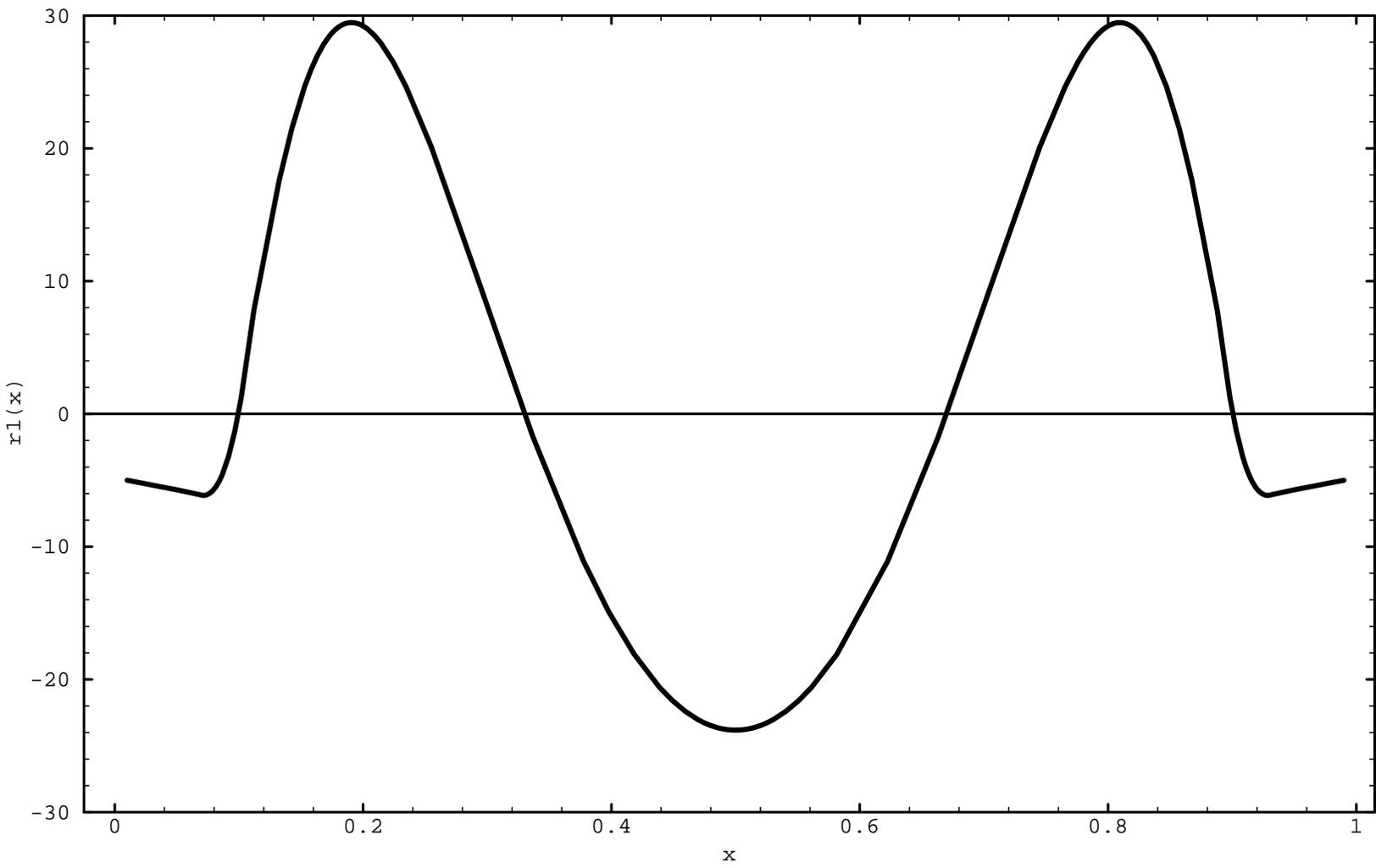

  Applying the algebraic procedures to the IT of SR
(\ref{ar-sr}) we arrive at the following results:
\begin{enumerate}
\item
For $\omega = \omega_{*} = 1.8 \,\mbox{GeV}^2$, the difference
$D_1\Phi_1(\omega_{*},1/2)$ is equal to zero and with increasing
$\omega$ it becomes negative (this means that $\varphi_1(1/2)<0$).
This value of the first-resonance mass is quite reasonable for
the pion case, the experimental one being
$\mu_1^{exp} = m_{\pi'}^2 \approx 1.7 \ \mbox{GeV}^2$
(at the full width $\Gamma_{\pi'}=0.2 \div 0.6 \ \mbox{GeV}$).
The pion WF $\varphi(x)$, corresponding to this $\omega_{*}$
has been obtained from expression (\ref{9}) in subsection 2.2
(see Fig. {fig-A2-123}). Its width is traditionally characterized
by the value of the second moment --
$\langle\xi^2\rangle_{\pi} \approx 0.17$. Another characteristic
often appearing in form factor calculations is the integral
$\int_{0}^{1} \varphi(x)/x dx \approx 2.8$.
Note that the method does not work properly in the neighborhood
of the endpoints $x=0$ and $x=1$ of the WF.
Therefore, to estimate the above values, one should exclude
these regions in the integral.

It should be emphasized that $\Phi_1(\omega,x)$ imitates the
l.h.s. of the SR as a function of $\omega$ rather well,
despite its different origin. Note here that the position of
the root $\omega_{*}$ depends also on the value of $x$:
in the region $0.35\le x\le 0.65$ this dependence is rather weak.
We don't see the step structure on the l.h.s. of Eq. (\ref{eq:D1})
-- the structure is destroyed due to the employment
of too crude approximations in the theoretical part of the SR.
Following our experience in the "exactly solvable toy model"
considered in subsection 3.3, we may hope,  nevertheless, that
the roots of the few trigger differences
$D_{1,2, \ldots} \Phi_N(\omega_{*},1/2)$
yield reliable approximations for the masses.

\item
The saturation by the first resonance is reached near
$\omega \approx  3 ~\mbox{GeV}^2$. The curves correspon\-ding to
the resonance $\varphi_1(x)$ (for $\omega \approx 3~\mbox{GeV}^2$)
are shown in Fig. \ref{fig-A2-res}.

\item
The last information we can extract in this case is the position
of the second resonance: the equation $D_2\Phi_1(\omega_{**},x)=0$
gives us the range of values
$\omega_{**}= \mu_2 = 3 \div 4 \mbox{GeV}^2$, which is
converted to the mass value $m_2 = 1.86\pm 0.14 \mbox{GeV}$.
The experimental parameters for this resonance are still
in question in the Particle Data booklet (April 1994).
However, the recent measurements in ~\cite{pi"95}
certainly provide for the mass $m_{\pi''} \approx 1.78 \pm 0.007 \
\mbox{GeV}$ (at full width
$\Gamma_{\pi''} = 0.16 \ \mbox{GeV}$), which is rather close
to our estimate.
\end{enumerate}

\section {Conclusion}
We have considered a model SR for the pion and the pseudoscalar
resonance WFs based on the non-diagonal correlator introduced
in \cite{arnew}. The theoretical side (r.h.s.) of this sum rule
depends only on the non-local condensate. We have tested
the approach proposed in our recent paper ~\cite{bm_wfp},
which enables one to extract WFs and masses from this SR,
using the exactly solvable toy model as an example.
We demonstrated the validity of the method and suggested
a pure algebraic procedure for extracting the masses
and WFs relating to the case under investigation.
We investigated also the stability of the procedure under
perturbations of the theoretical part of the sum rule.
We obtained as a result that the most crucial domain
in the complex plane of $\tau=M^{-2}$ is the neighborhood
of the poles lying on the real axes -- the perturbations
near these points essentially reorganize the spectral
expansion under investigation. Applying this method to
the pion case and using one of the ansatze given in
~\cite{bm_wfp} not only the mass and WF of the first resonance
($\pi'$) have been derived, but also the mass of $\pi''$.
Our results confirm the main conclusions about the shapes
of the WF of the pion and its resonances, obtained in
~\cite{arnew}.

 It should be mentioned that the specific form of the ansatz
for the correlation function $f(\nu)$ plays an important role
in determining concrete values of masses and shapes of the WFs
corresponding exactly to the pion resonances (on the contrary,
the WF of the pion ground state is not too sensitive to the
shape of the ansatz). We have chosen our ansatz (\ref{ansatz})
among others (see e.g.~\cite{bm_wfp}) just due to this reason.
So, one may consider it as a first step towards solving
the inverse problem, namely to obtain the vacuum condensate
properties from hadron phenomenology.
\vspace{15mm}

{\large \bf Acknowledgments}\\ \vspace{2mm}

We are grateful to ~A.~V.~Radyushkin for clarifying
criticism. We also thank  R.~Ruskov, A.~Dorokhov,
O.~Teryaev, N.~Stefanis and P.~Zhidkov for fruitful
discussions of the main results.
A special gratitude we express to N.~Stefanis for careful
reading the manuscript and giving many useful notes.
We are indebted to the International Science Foundation
(grant RFE-300), the Russian Foundation for Fundamental Research
(contract 93-02-3811) and the Heisenberg--Landau Program for
financial support.
We are grateful to Prof. H.~G.~Dosch for fruitful discussions
and warm hospitality at the Institute for Theoretical Physics
at Heidelberg University.

\newpage
\vspace*{80mm}

\begin{figure}[h]
  \caption{Forms of the WF of pion $\varphi_\pi(x)$ extracted from
           $\Phi_N(\omega\approx 1.5,x)$
           for $N=1$(line), $N=2$(dashed) and $N=3$(dotted)}
     \label{fig-A2-123}
      \end{figure}
\vspace{80mm}

\begin{figure}[h]
  \caption{Step-like structure of $\Phi_1(\omega,x)$ with
    $x = \mbox{fixed} = 0.5$ for the toy model case.}
     \label{fig-step}
      \end{figure}

\newpage
\vspace*{80mm}

\begin{figure}[h]
  \caption{Difference $D_1\Phi_1(\omega,x)$ with
     $x = \mbox{fixed} = 0.5$ for the exact r.h.s. of SR
     for the toy model case.}
     \label{fig-D1_All}
      \end{figure}
\vspace{80mm}

\begin{figure}[h]
  \caption{Difference $D_1\Phi_1(\omega,x,K=6)$ with
    $x = \mbox{fixed} = 0.5$  for the deformed $(K=6)$ r.h.s.
    of the SR for the toy model case.}
     \label{fig-D1_K6}
      \end{figure}
\newpage
\vspace*{80mm}

\begin{figure}[h]
  \caption{Difference $D_1\Phi_1(\omega,x,K=3)$
   with $x = \mbox{fixed} = 0.5$  for the deformed $(K=3)$ r.h.s.
   of the SR in the toy model case.}
     \label{fig-D1_K3}
      \end{figure}
\vspace{80mm}

\begin{figure}[h]
  \caption{Difference $D_1\Phi_1(\omega,x,K=6+1)$
  with $x = \mbox{fixed} = 0.5$ for the deformed $(K=6+1)$ r.h.s.
  of the SR in the toy model case.}
     \label{fig-D1_K61}
      \end{figure}
\newpage
\vspace*{80mm}

\begin{figure}[h]
  \caption{The deficit of the initial SR
  $\Delta_{SR}\{\Phi(\tau,x;K=6+1)\}$
  for the new spectrum $\mu_k=k$ in the toy model case.}
     \label{Def-SR-K=61_new}
      \end{figure}
\vspace{80mm}

\begin{figure}[h]
  \caption{The deficit of the initial SR
  $\Delta_{SR}\{\Phi(\tau,x;K=6+1)\}$ for the old spectrum
  $\mu_k=2k$ for the toy model case.}
     \label{Def-SR-K=61_old}
      \end{figure}
\newpage
\vspace*{80mm}

\begin{figure}[h]
  \caption{The difference between $\varphi_i(x)$
  and the exact WF of the $i$-th resonance for the $i=1, 2$
  for the deformed $(K=6)$ toy model case.}
     \label{fig-WF_res12}
      \end{figure}
\vspace{80mm}

\end{document}